\documentclass[prb,10pt,twocolumn]{revtex4}

\usepackage{epsfig}
\graphicspath{{/}}

\begin{document}

\title{Carrier Recombination and Generation Rates for Intravalley and Intervalley Phonon Scattering in Graphene}

\author{Farhan Rana, Paul A. George, Jared H. Strait, Jahan Dawlaty, Shriram Shivaraman, Mvs Chandrashekhar, Michael G. Spencer}
\affiliation{School of Electrical and Computer Engineering, Cornell University, Ithaca, NY 14853}
\email{fr37@cornell.edu}

\begin{abstract}
Electron-hole generation and recombination rates for intravalley and intervalley phonon scattering in Graphene are presented. The transverse and the longitudinal optical phonon modes ($E_{2g}$-modes) near the zone center ($\Gamma$-point) contribute to intravalley interband carrier scattering. At the zone edge ($K(K')$-point), only the transverse optical phonon mode ($A'_{1}$-mode) contributes significantly to intervalley interband scattering with recombination rates faster than those due to zone center phonons. The calculated recombination times range from less than a picosecond to more than hundreds of picoseconds and are strong functions of temperature and electron and hole densities. The theoretical calculations agree well with experimental measurements of the recombination rates of photoexcited carriers in graphene. 
\end{abstract}

\maketitle

\section{Introduction}
Graphene is a single two-dimensional atomic layer of carbon atoms forming a dense honeycomb crystal lattice~\cite{nov1,nov2,dressel1,heer}. The high mobility of electrons and holes in Graphene has prompted theoretical and experimental investigations into Graphene based ultra high speed electronic devices such as field-effect transistors, pn-junction diodes, and terahertz oscillators~\cite{shepard,marcus,rana,ryzhii}. The performance of many of these devices depends on the electron-hole generation and recombination rates in Graphene. For example, the diffusion length of injected minority carriers in a pn-junction diode is proportional to the square-root of the minority carrier recombination time. The threshold pumping levels required to achieve population inversion, and plasmon gain, in graphene also depend on the carrier recombination rates~\cite{rana}. Graphene photodetectors have also been realized~\cite{avouris}. The efficiency of graphene based visible/IR/far-IR detectors would also be critically dependent on the carrier recombination times~\cite{ryzhii2}. It is therefore important to understand the mechanisms that are responsible for electron-hole generation and recombination in graphene and the associated time scales.  

Intraband scattering  in graphene due to acoustic and optical phonons has been extensively studied~\cite{dressel2,ando1,ando2,sarma,mahan}. In this paper we calculate the electron-hole generation and recombination rates in graphene due to intravalley and intervalley phonon scattering. We find that the total recombination times range from less than a picosecond to more than hundreds of picoseconds and are strong functions of the temperature and the electron and hole densities. Near the zone center ($\Gamma$-point), only the transverse and the longitudinal optical phonon modes ($E_{2g}$-modes) contribute to intravalley interband scattering. At the zone edge ($K(K')$-point), only the transverse optical phonon mode ($A'_{1}$-mode) contributes significantly to intervalley interband carrier scattering and the calculated recombination rates are found to be faster than those due to the zone center $E_{2g}$-modes.  The theoretical results compare well (within a factor of unity) with the recently reported measurements of the recombination times of photoexcited carriers in epitaxial graphene~\cite{rana3}.

\section{Theoretical Model}
\begin{figure}[tbp]
  \begin{center}
   \epsfig{file=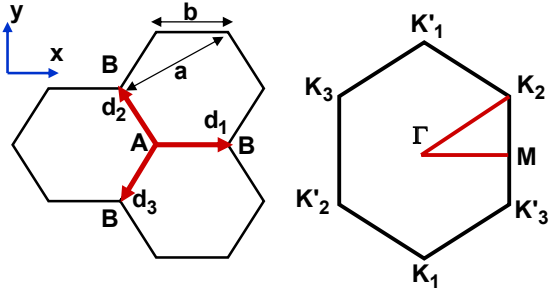,angle=-0,width=3.0 in}    
    \caption{LEFT: Graphene lattice. RIGHT: First Brillouin zone.}
    \label{fig1}
  \end{center}
\end{figure}

\begin{figure}[tbp]
  \begin{center}
   \epsfig{file=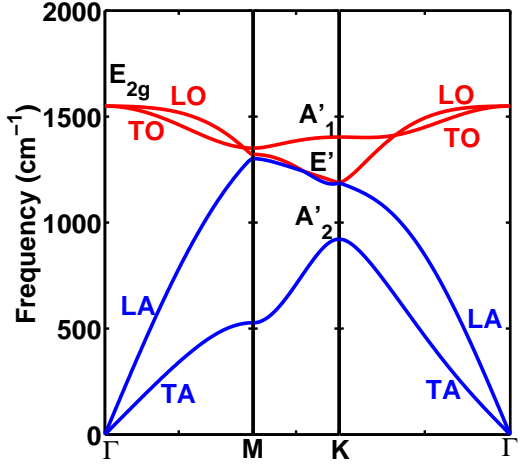,angle=-0,width=2.75 in}    
    \caption{Phonon dispersion relation of graphene obtained from considering fourth nearest neighbor interactions in the dynamical equation.}
    \label{fig2}
  \end{center}
\end{figure}

In graphene, the electronic hamiltonian for wavevector $\vec{k}$ in the tight binding approximation is~\cite{dressel1},
 \begin{equation}
H = \left[ \begin{array}{cc} 0 & t \, f(\vec{k}) \\
 t \,   f^{*}(\vec{k}) & 0  \end{array} \right] \label{eqH}
\end{equation}
where $t$ is the energy matrix element between the p-orbitals on neighboring carbon $A$ and $B$ atoms and has a value of $\sim$3.0 eV, and $f(\vec{k})=\sum_{m} \exp(i\,\vec{k}.\vec{d}_{m})$. Here, $\vec{d}_{m}$ are the position vectors from carbon atom $A$ to its three nearest $B$ atoms (Fig.\ref{fig1}). Near the $K$ (or $K'$) point in the Brillouin zone, the energy dispersion relation for the conduction and valence bands is, $E_{s}(\vec{k})=s\hbar\,v\,|\Delta \vec{k}|$. $s$ equals +1 and -1 for conduction and valence bands, respectively, the velocity $v$ equals $\sqrt{3} \, t\, a/2 = 10^{8}$ cm/s, and $\Delta \vec{k}$ is the difference between $\vec{k}$ and the wavevector of the $K$ (or $K'$) point. The corresponding Bloch function can be written as a sum over the p-orbitals of all carbon atoms, 
\begin{equation}
\psi_{\vec{k},s}(\vec{r})= \sum_{j} \frac{ e^{\displaystyle i\,\vec{k}.\vec{R}_{j}}}  {\sqrt{N}} \, c_{j}(\vec{k},s) \, p(\vec{r}-\vec{R}_{j})
\end{equation}   
For the $A$-atoms, $c_{j}(\vec{k},s)$ equals 1, and for the $B$-atoms, it equals $s f^{*}(\vec{k})/|f(\vec{k})|=s \exp(i\,\phi(\vec(k)))$. A suspended graphene sheet supports phonon modes with both in-plane and out-of-plane atomic displacements. The modes with out-of-plane atomic displacements do not couple with electronic transitions (to linear order in atomic displacements)~\cite{dressel2} and will therefore be ignored in this paper. The displacements $\vec{u}_{A}(\vec{q})$ and $\vec{u}_{B}(\vec{q})$ of the $A$ and $B$ atoms, respectively, corresponding to phonons with wavevector $\vec{q}$ satisfy the dynamical equation~\cite{dressel1}, 
\begin{equation}
\omega^{2}(\vec{k}) \, M \, \  \left[ \begin{array}{c} \vec{u}_{A}(\vec{q}) \\    \vec{u}_{B}(\vec{q}) \end{array} \right] = \overline{\overline{D}} (\vec{q}) \, \left[ \begin{array}{c} \vec{u}_{A}(\vec{q}) \\    \vec{u}_{B}(\vec{q}) \end{array} \right] \label{eqD}
\end{equation}
where, following Saito et. al.~\cite{dressel1}, the matrix $\overline{\overline{D}}$ includes fourth nearest neighbor interactions caused by in-plane bond-stretching and bond-bending displacements. $M$ is the mass of a carbon atom and $\omega(\vec{q})$ is the phonon frequency. The phonon dispersion relations found by solving the above equation are shown in Fig.~\ref{fig2}. Although this technique is not as accurate as the ab initio methods for obtaining phonon dispersions~\cite{abinitio}, it is adequate for calculating phonon eigenvectors at the high symmetry points and these eigenvectors will be needed in the calculations that follow. The LA and TA acoustic phonon modes near the $\Gamma$-point can cause intraband carrier transitions via deformation potential scattering~\cite{sarma2}. Zone center acoustic cannot cause interband carrier transitions because energy and momentum would not be conserved. Therefore we concentrate on the optical phonon modes. In the tight-binding approximation, the matrix element between two Bloch states of the perturbing hamiltonian due to atomic displacements associated with a phonon of wavevector $\vec{q}$ can be written as,  
\begin{eqnarray}
& & <\psi_{\vec{k'},s'}(\vec{r})| \Delta H |\psi_{\vec{k},s}(\vec{r}) > = \delta_{\vec{k}+\vec{q},\vec{k'}} \, \, \frac{\displaystyle 1}{\displaystyle 2} \, \frac{\displaystyle \partial \, t}{\displaystyle \partial b}  \nonumber \\
& & \times \, \sum_{m}  \frac{\displaystyle \vec{d}_{m}}{\displaystyle b} . \left[ \vec{u}_{B}(\vec{q}) \, e^{\displaystyle i\, \vec{q} . \vec{d}_{m}} - \vec{u}_{A}(\vec{q}) \right]  \nonumber \\
& & \times \, \left[ \, s' \, e^{\displaystyle - i\,\phi(\vec{k'})} \, e^{\displaystyle - i\,\vec{k'}.\vec{d}_{m}} +  s \, e^{\displaystyle i\,\phi(\vec{k})} \, e^{\displaystyle i\,\vec{k}.\vec{d}_{m}} \, \right] \label{eqmat}
\end{eqnarray}
Note that the electron wavevector $\vec{k}$ is measured form the zone center and the relation in (\ref{eqmat}) is valid for both intravalley and intervalley scattering processes. According to the approximation made here, only those phonon modes that cause bond-stretching couple with electronic transitions. The value of $\partial t/\partial b$ obtained from experiments and density functional calculations is $\sim$45 eV/nm~\cite{mauri}. The atomic displacements $\vec{u}_{A}(\vec{r})$ and $\vec{u}_{B}(\vec{r})$ for each phonon mode can be written in the standard form,
\begin{eqnarray}
\left[ \begin{array}{c} \vec{u}_{A}(\vec{r}) \\    \vec{u}_{B}(\vec{r}) \end{array} \right] = \sum_{\vec{q}} \, \sqrt{\frac{\hbar}{2 \, \rho \, \omega(\vec{q})}} \, \left[ \begin{array}{c} \vec{e}_{A}(\vec{q}) \\    \vec{e}_{B}(\vec{q}) \end{array} \right] & & \nonumber \\
\times \, \left( \, b_{\displaystyle \vec{q}} +  b^{*}_{\displaystyle -\vec{q}} \right) \,  \frac{e^{\displaystyle i\, \vec{q} . \vec{r} }}{\sqrt{V}} & & 
\end{eqnarray}
$\rho$ is the density of graphene and equals $\sim 7.6 \times 10^{-7}$ kg/m$^{2}$. It remains to calculate the eigenvectors $\vec{e}_{A}(\vec{q})$ and $\vec{e}_{B}(\vec{q})$ for phonon modes near the $\Gamma$ and $K\,(K')$ points. These can be obtained from the solution of the dynamical equation in (\ref{eqD}) and must be normalized such that, $\vec{e}^{*}_{A}(\vec{q}).\vec{e}_{A}(\vec{q})+\vec{e}^{*}_{B}(\vec{q}).\vec{e}_{B}(\vec{q}) = 2$. Near the $\Gamma$-point, the degenerate LO and TO phonon modes correspond to the two-dimensional $E_{2g}$ representation of the point group $D_{6h}$ and their eigenvectors can be written as,
\begin{equation}
 \left[ \begin{array}{c} \vec{e}_{A}(\vec{q}) \\    \vec{e}_{B}(\vec{q}) \end{array} \right] = \frac{1}{|\vec{q}|} \left[ \begin{array}{c} q_{x} \\ q_{y} \\ -q_{x} \\ -q_{y} \end{array} \right]
\,\,\,\,
 \left[ \begin{array}{c} \vec{e}_{A}(\vec{q}) \\    \vec{e}_{B}(\vec{q}) \end{array} \right] =  \frac{1}{|\vec{q}|} \left[ \begin{array}{c} -q_{y} \\ q_{x} \\ q_{y} \\ -q_{x} \end{array} \right]
\end{equation}
At the $K\,(K')$-points, the group of the wavevector is $D_{3h}$ and the phonon mode with the highest energy corresponds to the one-dimensional $A'_{1}$ representation of this group. At the $K_{1}$-point $4\pi/3a\,(0,-1)$, the eigenvetor is found from (\ref{eqD}) to be,
\begin{equation}
\left[ \begin{array}{c} \vec{e}_{A}(\vec{K_{1}}) \\ \vec{e}_{B}(\vec{K_{1}}) \end{array} \right] = \frac{1}{\sqrt{2}} \, \left[ \begin{array}{c} 1 \\ i \\ -1 \\ i \end{array} \right]
\end{equation}
The two degenerate modes at the $K\,(K')$-points correspond to the two-dimensional $E'$ representation of the group $D_{3h}$, and their eigenvectors at the $K_{1}$-point are,
\begin{equation}
\left[ \begin{array}{c} \vec{e}_{A}(\vec{K_{1}}) \\ \vec{e}_{B}(\vec{K_{1}}) \end{array} \right] = \frac{1}{\sqrt{2}} \, \left[ \begin{array}{c} 1 \\ -i \\ 1 \\ i \end{array} \right]
\,\,\,\,
 \left[ \begin{array}{c} \vec{e}_{A}(\vec{K_{1}}) \\ \vec{e}_{B}(\vec{K_{1}}) \end{array} \right] = \frac{1}{\sqrt{2}} \, \left[ \begin{array}{c} 1 \\ -i \\ -1 \\ -i \end{array} \right]
\end{equation}
Finally, the lowest energy phonon mode at the $K\,(K')$-points correspond to the one-dimensional $A'_{2}$ representation of the group $D_{3h}$ and its eigenvector at the $K_{1}$-point is,
\begin{equation}
 \left[ \begin{array}{c} \vec{e}_{A}(\vec{K_{1}}) \\ \vec{e}_{B}(\vec{K_{1}}) \end{array} \right] = \frac{1}{\sqrt{2}} \, \left[ \begin{array}{c} 1 \\ i \\ 1 \\ -i \end{array} \right]
\end{equation}
The eigenvectors for all other $K\,(K')$-points can be obtained either directly from (\ref{eqD}) or from the eigenvectors at the $K_{1}$-point using symmetry arguments based on the corresponding representation of the group $D_{3h}$. Using the calculated eigenvectors the matrix elements in (\ref{eqmat}) come out to be exactly zero for the $E'$ and the $A'_{2}$ phonon modes at the zone edge. Therefore, within the tight-binding approximation that only bond-stretching motion causes electron scattering, the $E'$ and the $A'_{2}$ phonon modes do not cause electron transitions. Away from the zone-edge, the matrix elements in (\ref{eqmat}) are non-zero for the $E'$ and the $A'_{2}$ modes. However, the net contribution to electron scattering is small enough compared to the other phonon modes that it will be ignored.  
\begin{figure}[tbp]
  \begin{center}
   \epsfig{file=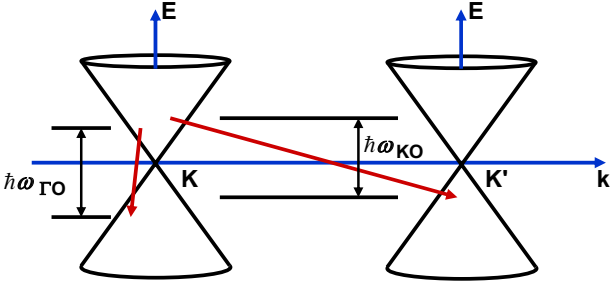,angle=-0,width=3.25 in}    
    \caption{Interband intravalley and intervalley electron scattering by optical phonons in graphene.}
    \label{fig3}
  \end{center}
\end{figure}

\section{Results}
Intravalley interband carrier scattering is due to the zone center LO and TO phonons ($E_{2g}$-modes). Assuming the frequency of these modes to be dispersion free and equal to $\omega_{\rm \Gamma O}$, the recombination rates $R_{\rm \Gamma LO}$ and $R_{\rm \Gamma TO}$, and generation rates $G_{\rm \Gamma LO}$ and $G_{\rm \Gamma LO}$ (units: 1/cm$^{2}$-s) can be obtained using the matrix elements given in (\ref{eqmat}) and the final results can be written as,
\begin{eqnarray}
& & R_{\rm \Gamma LO} =   R_{\rm \Gamma TO} =  \frac{9}{4}\,\left( \frac{\partial t}{\partial b} \right)^{2} \, \frac{1}{\displaystyle \pi \, \rho \, \omega_{\rm \Gamma O} \, \hbar^{4} \, v^{4}} \, \int_{0}^{\hbar \omega_{\rm \Gamma 0}}  dE  \nonumber \\
& & \times \, E \, \left( \hbar \omega_{\rm \Gamma O} - E \right) \, f(E-E_{fc}) \nonumber \\
& & \times \, \left[1-f(E-\hbar\omega_{\rm \Gamma O}-E_{fv})\right]\,\left[ 1 + n_{ph}(\hbar \omega_{\rm \Gamma O}) \right] \label{eqr1}
\end{eqnarray} 
\begin{eqnarray}
& & G_{\rm \Gamma LO} =   G_{\rm \Gamma TO} =  \frac{9}{4}\,\left( \frac{\partial t}{\partial b} \right)^{2} \, \frac{1}{\displaystyle \pi \, \rho \, \omega_{\rm \Gamma O} \, \hbar^{4} \, v^{4}} \, \int_{0}^{\hbar \omega_{\Gamma 0}}  dE  \nonumber \\
& & \times \, E \, \left( \hbar \omega_{\rm \Gamma O} - E) \right) \, f(E-\hbar\omega_{\rm \Gamma O}-E_{fv}) \nonumber \\
& & \times \, \left[1-f(E-E_{fc})\right]\,n_{ph}(\hbar \omega_{\rm \Gamma O}) \label{eqr2}
\end{eqnarray} 
Here, $f(E-E_{f})$ are the electron distribution functions for the conduction and valence bands and $n_{ph}(\hbar \omega_{\rm \Gamma O})$ are the phonon occupation numbers. The product $E(\hbar \omega_{\rm \Gamma O} - E)$ in the integrand comes from the initial and final density of states. Intervalley interband scattering is due to the zone edge $A'_{1}$-mode. Assuming the frequency of this mode to be dispersion free and equal to $\omega_{\rm K O}$, the recombination and generation rates, $R_{\rm K O}$ and $G_{\rm K O}$ (units: 1/cm$^{2}$-s), respectively, can be obtained in a similar fashion. The results are identical to those given in (\ref{eqr1}) and (\ref{eqr2}) except that the prefactors are $9/2$ instead of $9/4$ (i.e. twice as large), and $\omega_{\rm \Gamma O}$ is replaced by $\omega_{\rm K O}$. The total recombination rate $R$ and the generation rate $G$  due to both intravalley and intervalley phonon scattering are, $R = R_{\rm \Gamma LO} + R_{\rm \Gamma TO} + R_{\rm K O}$ and $G= G_{\rm \Gamma LO} + G_{\rm \Gamma TO} + G_{\rm K O}$. Since $\omega_{\rm K O}<\omega_{\rm \Gamma O}$ ($\omega_{\rm K O} \sim 1300$ cm$^{-1}$ and $\omega_{\rm \Gamma O} \sim 1580$ cm$^{-1}$), and because the integrated squared matrix element for scattering by the zone edge $A'_{1}$-mode is twice as large as the one for each of the zone center $E_{2g}$-modes, electron-hole recombination due to intervalley scattering is expected to be faster than due to intravalley scattering. 

It is interesting to look at the energy dependence of the recombination lifetime $\tau_{r}(E)$ of an electron in the conduction band. Assuming $n_{ph} \approx 0$, $\tau_{r}(E)$ is,
\begin{eqnarray}
\frac{1}{\displaystyle \tau_{r}(E)} & = & \frac{9}{4}\,\left( \frac{\partial t}{\partial b} \right)^{2} \, \frac{\left( \hbar \omega_{\rm \Gamma O} - E \right)}{\displaystyle \rho \, \omega_{\rm \Gamma O} \, \hbar^{2} \, v^{2}} \, \left[1-f(E-\hbar\omega_{\rm \Gamma O}-E_{fv})\right] \nonumber \\
& + & \left\{ \omega_{\rm \Gamma O} \rightarrow \omega_{\rm K O} \right\} \label{eqml}
\end{eqnarray}
The above expression shows that if the valence band has enough empty states available then conduction electrons near the Dirac point recombine the fastest since they see the largest final density of states in the valence band. If the valence band has empty states only near the Dirac point then conduction electrons with energy near the optical phonon energy will have the shortest recombination lifetime. The average recombination time $\tau_{r}$ is defined as,
\begin{equation}
\tau_{r} = \frac{{\rm min}(n,p)}{R}
\end{equation}
\begin{figure}[tbp]
  \begin{center}
   \epsfig{file=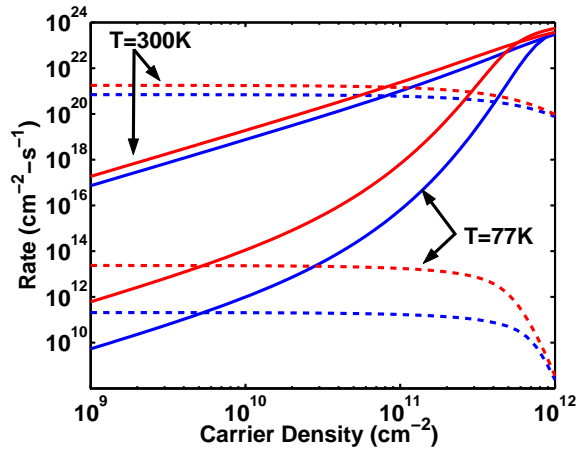,angle=0,width=3.0 in}    
    \caption{Electron-hole recombination (SOLID) and generation (DASHED) rates due to intravalley and intervalley optical phonon scattering in graphene at T=77K and at T=300K are plotted as a function of the electron and hole density (assumed to be equal). The top curve in each pair corresponds to intervalley scattering ($R_{\rm KO}$ or $G_{\rm K O}$) and the bottom curve corresponds to intravalley scattering ($R_{\rm \Gamma LO}+R_{\rm \Gamma TO}$ or $G_{\rm \Gamma LO}+G_{\rm \Gamma TO}$).}   
    \label{fig4}
  \end{center}
\end{figure}
\begin{figure}[tbp]
  \begin{center}
   \epsfig{file=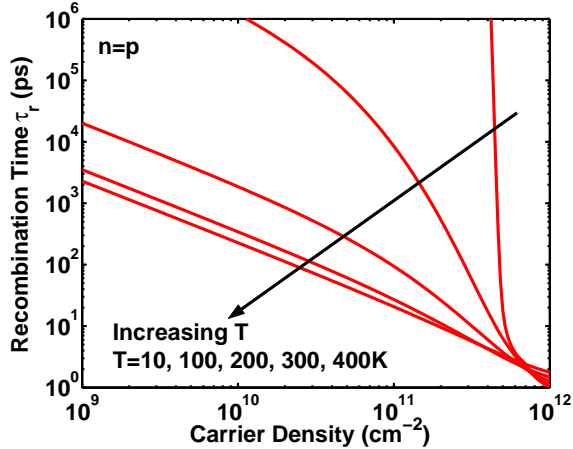,angle=0,width=3.0 in}    
    \caption{Electron-hole recombination time $\tau_{r}$ due to intravalley and intervalley optical phonon scattering in graphene is plotted as a function of the electron and hole density (assumed to be equal) for different temperatures.}   
    \label{fig5}
  \end{center}
\end{figure}
\begin{figure}[tbp]
  \begin{center}
   \epsfig{file=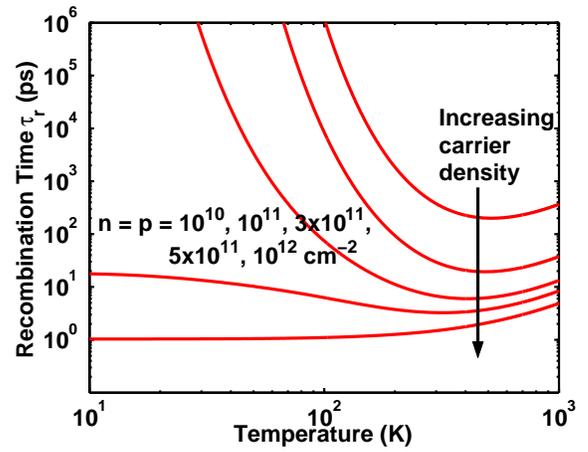,angle=0,width=3.0 in}    
    \caption{Electron-hole recombination time $\tau_{r}$ due to intravalley and intervalley optical phonon scattering in graphene is plotted as a function of the temperature for different electron and hole densities (assumed to be equal).}   
    \label{fig6}
  \end{center}
\end{figure}

Fig.~\ref{fig4} plots the recombination and generation rates due to intravalley and intervalley optical phonon scattering at T=77K and at T=300K are plotted as a function of the electron and hole density (assumed to be equal). Equal electron and hole densities are generated, for example, when graphene is photoexcited as was done in recent experiments~\cite{avouris,ryzhii2,rana3,rana4,norris}. Near thermal equilibrium electron and hole densities, recombination and generation rates due to intervalley scattering ($R_{\rm KO}$ or $G_{\rm K O}$) are $\sim$2 times faster at T=300K and $\sim$50 times faster at T=77K compared to intravalley scattering rates ($R_{\rm \Gamma LO}+R_{\rm \Gamma TO}$ or $G_{\rm \Gamma LO}+G_{\rm \Gamma TO}$). This difference is because $\omega_{\rm K O}<\omega_{\rm \Gamma O}$. As would be expected, interband optical phonon scattering is less effective at lower temperatures and at smaller carrier densities. However, even at low temperatures the scattering rates can be very fast if the carrier densities are large. Fig.~\ref{fig5} shows the average recombination time $\tau_{r}$ due to phonon scattering plotted as a function of the electron and hole density (assumed to be equal) for different temperatures. Fig.~\ref{fig6} shows the recombination time $\tau_{r}$ plotted as a function of the temperature for different carrier densities. It can be seen that for carrier densities larger than mid-$10^{11}$ cm$^{-2}$ the recombination times are shorter than 10 ps even at low temperatures. 
\begin{figure}[tbp]
  \begin{center}
   \epsfig{file=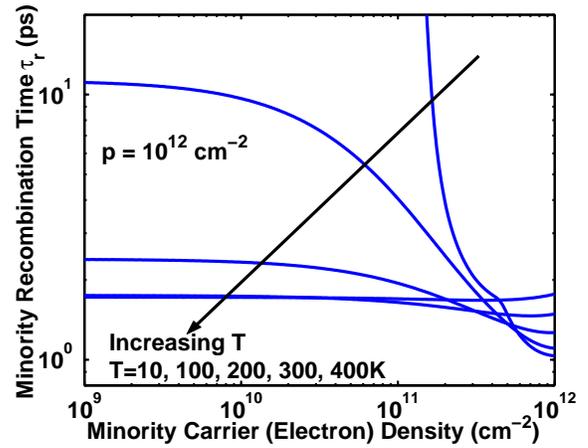,angle=0,width=3.0 in}    
    \caption{Minority carrier (electron) average recombination times due to intravalley and intervalley phonon scattering are plotted as a function of the minority carrier density for different temperatures. The majority carrier (hole) density is assumed to be $10^{12}$ cm$^{-2}$.}   
    \label{fig7}
  \end{center}
\end{figure}
\begin{figure}[tbp]
  \begin{center}
   \epsfig{file=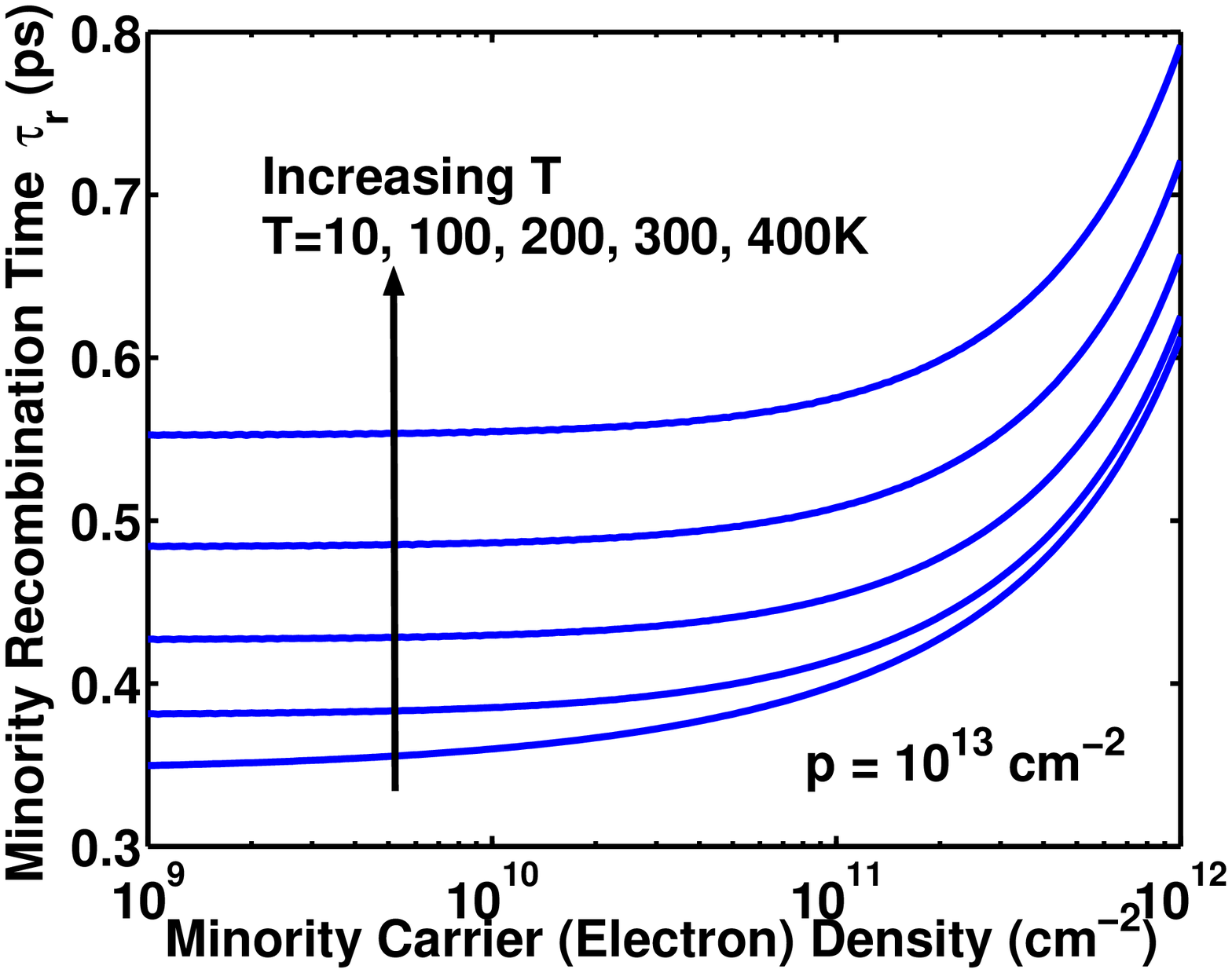,angle=0,width=3.0 in}    
    \caption{Minority carrier (electron) average recombination times due to intravalley and intervalley phonon scattering are plotted as a function of the minority carrier density for different temperatures. The majority carrier (hole) density is assumed to be $10^{13}$ cm$^{-2}$.}  
    \label{fig8}
  \end{center}
\end{figure}
\begin{figure}[tbp]
  \begin{center}
   \epsfig{file=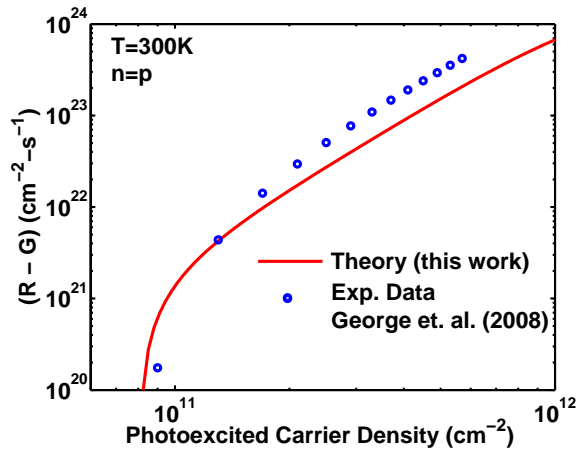,angle=0,width=3.0 in}    
    \caption{The calculated net recombination rate, defined as $R-G$, due to interband phonon scattering is plotted as a function of the electron and hole density (assumed to be equal) along with the experimental data from George et. al.~\cite{rana3}. T=300K.}  
    \label{fig9}
  \end{center}
\end{figure}

Another important quantity for device applications is the minority carrier recombination time in doped (chemically or electrostatically) graphene. Assuming electrons to be the minority carrier, Figs.~\ref{fig7} and ~\ref{fig8} plot the minority carrier average recombination times due to phonon scattering as a function of the minority carrier density for different temperatures. The majority carrier density is assumed to be $10^{12}$ cm$^{-2}$ (Fig.~\ref{fig7}) and $10^{13}$ cm$^{-2}$ (Fig.~\ref{fig8}). The temperature dependence of the minority carrier lifetime is seen to depend on the majority carrier density. For small majority carrier densities, the minority carriers near the Dirac point cannot recombine and therefore the minority carrier lifetime decreases with temperature. For large majority carrier densities, the minority carriers near the Dirac point can recombine very fast and therefore the minority carrier lifetime increases with temperature. Fig.~\ref{fig8} shows that the minority carrier recombination times can be shorter than a picosecond for highly doped graphene. For large majority carrier densities, the minority carrier lifetime for small minority carrier densities and low temperatures follows from (\ref{eqml}) and equals $\tau_{r}(E=0)$ ($\approx 0.34$ ps).

Recently, electron-hole recombination rates in graphene were measured using ultrafast optical-pump and terahertz-probe spectroscopy~\cite{rana3}. Fig.~\ref{fig9} shows a comparison of the experimental data with the present work. Fig.~\ref{fig9} plots the calculated net recombination rate, defined as $R-G$, due to interband phonon scattering as a function of the electron and hole density (assumed to be equal) at T=300K. Also shown are the best fits to the measured time-resolved data from George et. al~\cite{rana3}. The theory compares well (within a factor unity) with the experiments. The measured rates are between 2.0-2.5 times faster than the calculated rates. This difference is likely due to the exclusion of other recombination mechanisms in calculating the rates plotted in Fig.~\ref{fig9}. For example, Coulomb scattering (Auger recombination)~\cite{rana2} and plasmon emission~\cite{rana} can also contribute to electron-hole recombination. In particular, Auger recombination rates calculated by Rana~\cite{rana2} are of nearly the same magnitude as the phonon recombination rates shown in Fig.~\ref{fig9} for carrier densities in the $10^{11}$-$10^{12}$ cm$^{-2}$ range. In addition, the presence of disorder can also enhance recombination rates, as was experimentally observed by George et. al.~\cite{rana3}, by providing the additional momentum needed to satisfy conservation rules in interband scattering proceses which would otherwise be prohibited.    

\acknowledgements
The authors would like to acknowledge helpful discussions with Paul Mceuen and Jiwoong Park. The authors acknowledge support from Eric Johnson through the National Science Foundation, the DARPA Young Faculty Award, the Air Force Office of Scientific Research (Contract No. FA9550-07-1-0332, monitor Donald Silversmith), and the Cornell Material Science and Engineering Center (CCMR) program of the National Science Foundation (Cooperative Agreement No. 0520404).

\end{document}